\documentclass[12pt]{iopart}

\usepackage{iopams}
\usepackage{color}
\usepackage[dvips]{graphicx}

\begin{document}

\title[Spatiotemporal correlations in entangled pairs of photons]{Spatiotemporal correlations in entangled photons
\\generated by spontaneous parametric down conversion}

\author{Clara I. Osorio$^{1}$, Alejandra Valencia$^{1}$ and Juan P. Torres$^{1,2}$}
\address{$^{1}$ICFO-Institut de Ciencies Fotoniques,
Mediterranean Technology Park, 08860 Castelldefels (Barcelona),
Spain\\ $^{2}$Dept. Signal Theory and Communications, Universitat
Politecnica de Catalunya, Jordi Girona 1-3, 08034 Barcelona Spain}
\ead{clara.ines.osorio@icfo.es}

\begin{abstract}
In most configurations aimed at generating entangled photons based
on spontaneous parametric down conversion (SPDC), the generated
pairs of photons are required to be entangled in only one degree
of freedom. Any distinguishing information coming from the other
degrees of freedom that characterize the photon should be
suppressed to avoid correlations with the degree of freedom of
interest. However, this suppression is not always possible. Here,
we show how the frequency information available affects the purity
of the two-photon state in space, revealing a correlation between
the frequency and the space degrees of freedom. This correlation
should be taken into account to calculate the total amount of
entanglement between the photons.
\end{abstract}
 \maketitle

\section{Introduction}
Pairs of photons generated via spontaneous parametric down
conversion (SPDC) are widely used in quantum cryptography, to test
violation of Bell's inequalities and in other quantum information
tasks, including dense coding and teleportation \cite{nielsen1}.
The popularity of this source of paired photons is strongly
related to the relative simplicity of its experimental
realization, and to the variety of quantum features that down
converted photons can exhibit. It is well known that a pair of
photons generated via SPDC can be entangled in different degrees
of freedom, for example in polarization \cite{walborn2,kwiat}, in
frequency \cite{mosley,valencia} and in the equivalent degrees of
freedom: orbital angular momentum \cite{mair,molina-terriza},
space and transverse momentum \cite{walborn,law2}.\par

Usually, when considering entanglement between the two generated
photons, only one of the degrees of freedom in which the pair may
be entangled is considered and any influence of the others is
neglected. This is often justified by the assumption of a
filtering process in which the use of very narrow filters in one degree of freedom allows to neglect it
\cite{osorio}. Only one degree of freedom is considered, for
example, in the experimental demonstration of teleportation
\cite{bouwmeester} and in some proposals to measure the spatial
entanglement between paired photons \cite{walborn}.
\par

Pairs of photons generated with specific characteristics in
various degrees of freedom have been also considered. For example,
in hyperentanglement \cite{barreiro,barbieri} where the photons
are entangled in various degrees of freedom that are assumed to be
independent and, in configurations where one degree of freedom
modifies the quantum state of the photons in other degrees of
freedom:  to distillate position entanglement using the
polarization \cite{caetano} or to control the joint frequency
distribution of the generated pair by modifying the spatial
properties of the pump beam \cite{valencia}.

All the configurations mentioned before assume a specific
relationship between frequency, polarization and space in the
two-photon state. In this paper, we will concentrate on the
frequency and spatial properties of a pure two-photon state.  We
will show that the purity of the spatial two-photon state
generated by SPDC depends on the frequency information available.
This imply a correlation between space and frequency that cannot
be ignored in any real filtering process, where the finite
bandwidth of the filters affects the separability of the frequency
and the space degrees of freedom. This is analogous to the effect
of the spatial \cite{vanexter2} and frequency \cite{humble}
correlations of the paired photons on the degree of polarization
entanglement of the generated photons. Additionally, in this
letter we will show how the spatiotemporal correlations affect the
entanglement between the photon. We will calculate how the purity
of the signal photon in space and frequency, and therefore the
amount of entanglement between the photons, depends on the
geometry of the SPDC configuration, the characteristics of the
pump beam and the filtering process.

This paper is divided in three sections. In section \ref{secGC} we
discuss the main characteristics of the two-photon quantum state
generated in spontaneous parametric down conversion considering
the space and frequency degrees of freedom.  In section
\ref{svsf}, the spatial state of the two-photon is described. It
is shown how the spatial state may become mixed as a consequence
of tracing out the frequency variables, which implies correlation
between those degrees of freedom.  Finally, in section \ref{svsi},
the purity of the signal photon state is analyzed taking into
account the space and the frequency degrees of freedom.

\section{Quantum State of pairs of photons generated via spontaneous parametric down conversion}\label{secGC}

\subsection{General case}

Spontaneous parametric down conversion (SPDC) is an optical process in which a
nonlinear crystal of length $L$ is illuminated by a laser pump
beam propagating in the $z$ direction. Due to the interaction of
a pump photon  with the crystal there is a small probability to
generate a pair of photons: signal and idler. In general, the
direction of propagation of the pump beam is modified by a small
angle $\rho_0$ due to the presence of Poynting vector walk-off
inside the crystal. All photons that interacts in the nonlinear
process are characterized by the transverse momentum
$\mathbf{q_n}=(q_n^x,q_n^y)$,  and the frecuency
$\omega_n=\omega_n^0+\Omega_n$, where $\omega_n^0$ is the central
frequency and $\Omega_n$ is the frequency deviation. The index $n$
labels the signal $(n=s)$, idler $(n=i)$ and pump $(n=p)$.

For the sake of simplicity, in what follows we will assume that in
the frequency and space degrees of freedom, the two-photon state
is pure \cite{molina1}, therefore the density matrix reads

\begin{eqnarray}
 \rho =&& \int d\mathbf{q}_sd\Omega_sd\mathbf{q}_id\Omega_i
    d\mathbf{q}_s'd\Omega_s'd\mathbf{q}_i'd\Omega_i' \\\nonumber
&&\times\Phi(\mathbf{q}_s,\Omega_s,\mathbf{q}_i,\Omega_i)
    \Phi^{*}(\mathbf{q}'_s,\Omega'_s,\mathbf{q}'_i,\Omega'_i)\\\nonumber
    && \times|\mathbf{q}_s,\Omega_s\rangle|\mathbf{q}_i,\Omega_i\rangle\langle
    \mathbf{q}'_s,\Omega'_s|\langle\mathbf{q}'_i,\Omega'_i|. \\\nonumber
    \label{densitymatrix}
\end{eqnarray}

\noindent The spatiotemporal mode function
$\Phi(\mathbf{q}_s,\Omega_s,\mathbf{q}_i,\Omega_i)$ in a
noncollinear configuration is given by

\begin{eqnarray}     \label{modefunction}
\Phi(\mathbf{q}_s,\Omega_s,\mathbf{q}_i,\Omega_i) &&\propto
E_p(q_s^x+q_i^x,\Delta_0) B_p(\Omega_s+\Omega_i)\\\nonumber
&&\times\mathcal{C}_{spatial}(\mathbf{q}_s)\mathcal{C}_{spatial}(\mathbf{q}_i)\\\nonumber
&&\times\mathcal{F}_{frequency}(\Omega_s)\mathcal{F}_{frequency}(\Omega_i)\\\nonumber
&&\times\mathrm{sinc} \left(\frac{\Delta_k L}{2}\right)\\\nonumber
\end{eqnarray}

\noindent  where $B_p(\omega_p^0+\Omega_p)$ and
$E_p(\mathbf{q}_p)$ are the frequency and transverse momentum
distribution of the pump beam at the center of the nonlinear
crystal, respectively. The effect of the unavoidable spatial
filtering produced by the specific optical detection system is
described by the spatial collection function
$\mathcal{C}_{spatial}(\mathbf{q}_n)$. The presence of frequency
filters in the paths of signal and idler is indicated by frequency
filter function $\mathcal{F}_{frequency}(\Omega_n)$.
 The Phase matching conditions appears in
equation~(\ref{modefunction}) through  the ``Delta'' factors,
defined by

\begin{eqnarray} \label{deltas}
\Delta_0&=&q_s^y \cos\varphi_s+q_i^y \cos\varphi_i+k_s
\sin\varphi_s-k_i \sin\varphi_i\\\nonumber
 \Delta_k&=&k_p-k_s
\cos\varphi_s-k_i \cos\varphi_i-q_s^y \sin\varphi_s+q_i^y
\sin\varphi_i\\\nonumber &&+(q_s^x+q_i^x)\tan \rho_0\cos
\alpha+\Delta_0 \tan\rho_0\sin \alpha
\\\nonumber
\end{eqnarray}

\noindent where  $k_n=\left[ \left(\omega_n^0 n_n / c\right)^2
-|\mathbf{q}_n|^2
    \right]^{1/2}$ is the longitudinal wave vector inside the crystal.
$\varphi_{s}$ and $\varphi_{i}$ are the propagation directions of
the generated photons inside the crystal with respect to the pump
direction $z$ and $\alpha$ is the azimuthal angle
\cite{osorio,torres2}.

\subsection{Gaussian approximations}
In order to find analytical expressions for the purity of the
spatial two-photon state and the purity of the  signal state in
frequency and space, we make some assumptions and approximations
following refs \cite{joobeur,torres3}.

We consider a gaussian shape for the spatial and frequency
distributions of the pump beam:

\begin{equation}\label{gaussianspatial}
    E_p(q_p^x,q_p^y)=\exp\left[-
    \frac{\mathrm{w}_p^2}{4}(q_p^{x2}+q_p^{y2})
    \right]
\end{equation}

\noindent and

\begin{equation}\label{gaussianfrequency}
 B_p(\Omega_s+\Omega_i)= \exp\left[-
    \frac{T_0^2}{4}\Omega_p^2
    \right].
\end{equation}

\noindent The pump beam waist is given by $\mathrm{w}_p$ and the
pulse temporal  duration is $T_0$.

Both, the  spatial collection and the frequency filter functions
are described by Gaussians, so that $
\mathcal{C}_{spatial}(\mathbf{q}_n)\propto\exp\left[-\mathrm{w}_n^2|\mathbf{q}_n|^2/2\right]$
and
$\mathcal{F}_{frequency}(\Omega_n)\propto\exp\left[-\Omega_n^2/(2
B_n^2)\right]$ where $\mathrm{w}_n$  is the spatial collection
mode width and $B_n$ is the frequency bandwidth. Notice that, in
order to have more convenient units, the filters in momentum are
defined in a different way than the filters in frequency. While
$B_{s,i} \rightarrow 0$ implies a single frequency collection, the
condition for a single q vector collection is  $w_{s,i}
\rightarrow \infty$. In what follows we consider $B_s=B_i$ and
$\mathrm{w}_s=\mathrm{w}_i$.

Under these considerations and by making the approximation $
\mathrm{sinc} \left(\Delta_k L/2\right)\approx \exp
\left[-\beta^2\Delta_k^2 L^2/4\right]$ with $\beta=0.455$, the
mode function of the generated pair can be written as

\begin{equation}\label{exponentialMF}
    \Phi(\mathbf{q}_s,\Omega_s,\mathbf{q}_i,\Omega_i)= N\exp
    \left[-\frac{1}{2}x^tA x\right]
\end{equation}

\noindent where $N$ is a normalization constant, $A$ is a
positive-definite real $6\times6$ matrix containing all the
parameters that describes the SPDC process
and $x^t$ is the transpose of the vector $x$ given by $x^t=\left(%
\begin{array}{cccccc}
  q_s^x, & q_s^y, & \Omega_s, & q_i^x, & q_i^y, & \Omega_i \\
\end{array}%
\right)$.

\begin{figure}[thb]
\centering
\includegraphics[scale=0.50]{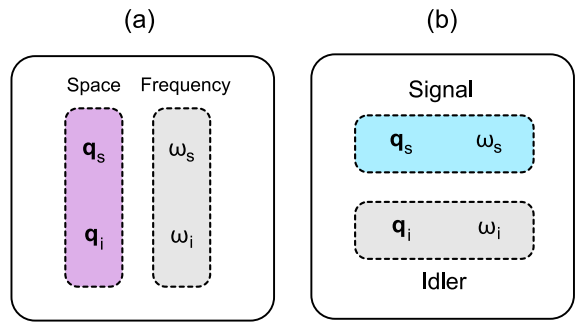}
\caption{The two-photon state in space and frequency can be
considered as composed by two subsystems. In (a) those subsystems
are the degrees of freedom. The frequency, in gray, is traced out
in the calculations of Sec. \ref{svsf}. In (b) the subsystems
considered are the generated photons. In gray we show the idler
photon that is traced out in the calculations of Sec. \ref{svsi}.}
\label{comparation}
\end{figure}

The physical system described by equation~(\ref{exponentialMF})
can be considered as composed by two subsystems. These  subsystems
can either be the two degrees of freedom, figure~\ref{comparation}
(a) or the two photons, figure~\ref{comparation} (b). In both
cases, the correlations that may exist between the subsystems
imply that the system is not in a separable state, i.e., when
considering the two degrees of freedom
$\Phi(\mathbf{q}_s,\Omega_s,\mathbf{q}_i,\Omega_i)
\neq\Phi_{\mathbf{q}}(\mathbf{q}_s,\mathbf{q}_i)
\Phi_{\Omega}(\Omega_s,\Omega_i)$ and when considering the two
photons $\Phi(\mathbf{q}_s,\Omega_s,\mathbf{q}_i,\Omega_i)
\neq\Phi_s(\mathbf{q}_s,\Omega_s) \Phi_i(\mathbf{q}_i,\Omega_i)$.

\section{Correlations between space and frequency degrees of freedom}\label{svsf}
Let us consider the two degrees of freedom as subsystems like in
figure ~\ref{comparation}(a). The state of one subsystem can be
calculated tracing out the parameters that describe the other
subsystem.  Tracing out the frequency, the reduced density matrix
for the resulting spatial two-photon state is given by

\begin{eqnarray}
    \rho_{\mathbf{q}} &=& Tr_{\Omega}(\rho)\\\nonumber
    &=& \int d \Omega_s''d\Omega_i'' \langle \Omega_s'',
    \Omega_i''|\rho|\Omega_s'' \Omega_i''\rangle\\\nonumber
        &=&\int d\mathbf{q}_sd\Omega_sd\mathbf{q}_id\Omega_i
    d\mathbf{q}_s'd\mathbf{q}_i'\\\nonumber
    &&\Phi(\mathbf{q}_s,\Omega_s,\mathbf{q}_i,\Omega_i)
    \Phi^{*}(\mathbf{q}'_s,\Omega_s,\mathbf{q}'_i,\Omega_i)
    |\mathbf{q}_s,\mathbf{q}_i\rangle\langle
    \mathbf{q}'_s,\mathbf{q}'_i|\\\nonumber
\label{spatialrho}
\end{eqnarray}

\noindent and the corresponding purity  $Tr(\rho_{\mathbf{q}}^2)$
of the spatial two-photon state is

\begin{eqnarray}
    Tr(\rho_{\mathbf{q}}^2)&&=
    \int d\mathbf{q}_sd\Omega_sd\mathbf{q}_id\Omega_i
    d\mathbf{q}_s'd\Omega_s'd\mathbf{q}_i'd\Omega_i'\\\nonumber
    &&\times\Phi(\mathbf{q}_s,\Omega_s,\mathbf{q}_i,\Omega_i)
    \Phi^{*}(\mathbf{q}'_s,\Omega_s,\mathbf{q}_i',\Omega_i)\\\nonumber
    &&\times\Phi(\mathbf{q}_s',\Omega_s',\mathbf{q}_i',\Omega_i')
    \Phi^{*}(\mathbf{q}_s,\Omega_s',\mathbf{q}_i,\Omega_i').
     \\\nonumber
     \label{puritydefspatial}
\end{eqnarray}

\noindent We can solve this integral using the exponential
character of the mode function given by
equation~(\ref{exponentialMF}), so that the purity becomes
\begin{equation}\label{purityfreq}
    Tr(\rho_{\mathbf{q}}^2)=\frac{\det(2A)}{\sqrt{\det(B)}}
\end{equation}

\noindent where  $B$ is a positive-definite real $12\times12$
matrix given by

\begin{eqnarray}\label{exponentialPF}
&&N^4\exp
    \left[-\frac{1}{2}X^tB
    X\right]=\Phi(\mathbf{q}_s,\Omega_s,\mathbf{q}_i,\Omega_i)\\\nonumber
 &&\times    \Phi^{*}(\mathbf{q}'_s,\Omega_s,\mathbf{q}'_i,\Omega_i)\Phi(\mathbf{q}_s',\Omega_s',\mathbf{q}_i',\Omega_i')
    \Phi^{*}(\mathbf{q}_s,\Omega_s',\mathbf{q}_i,\Omega_i').\\\nonumber
\end{eqnarray}

\noindent $X^t=\left( \begin{array}{cc}
 x^t, & x'^t\\\end{array}\right)$ is a vector resulting from the concatenation of
$x^t$ and $x'^t=\left(
\begin{array}{cccccc}
  q_s^{x'}, & q_s^{y'}, & \Omega_s', & q_i^{x'}, & q_i^{y'}, & \Omega'_i \\
\end{array}\right)$.

Since the composed quantum system described by
equation~(\ref{densitymatrix}) is in a pure state
($Tr(\rho^2)=1$), the purity of the spatial two-photon state given
by equation \ref{purityfreq}, is different from $1$ only if the
degrees of freedom are correlated \cite{nielsen}. This correlation
is stronger when the partial purity gets closer to $0$
\cite{nielsen1}.

Notice that the purity of the spatial two-photon state is equal to
the purity of the frequency two-photon state when the spatial
degree of freedom is traced out. Therefore all the results derived
for the spatial two-photon state can be extended to a frequency
two-photon state.

\begin{figure}[thb]
\centering
\includegraphics[bb= 60   227   526   670,scale=0.50]{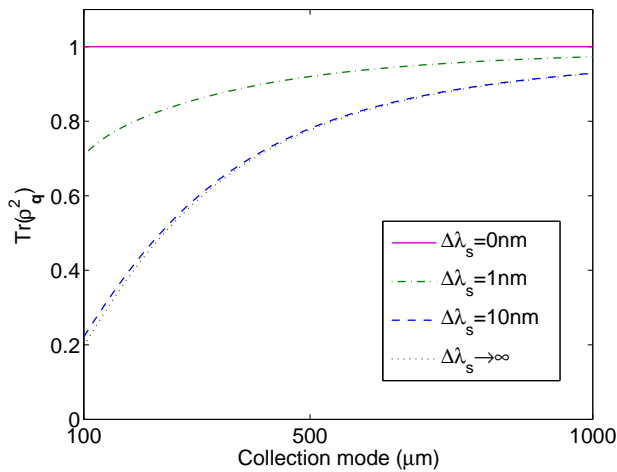}
\caption{Purity of the spatial state of the pair of generated
photons for different values of the frequency filter bandwidth, as
a function of the spatial collection mode width
($\mathrm{w}_{s}=\mathrm{w}_{i}$). The pump beam waist is
$\mathrm{w}_p=400\mu m$ and the emission angle is
$\varphi=10^{\circ}$.} \label{wsdeltas}
\end{figure}

We plot the purity of the spatial state $Tr(\rho_{\mathbf{q}}^2)$
as a function of the spatial filter width for different frequency
filter bandwidths, figure ~\ref{wsdeltas}.   The half width at
$1/e$ of the frequency filters in the wavelength variable is given
by $\Delta\lambda_n$, that is  related to $B_s$  and $B_i$ by
$B_{n}=\pi c \Delta\lambda_n/(\lambda_n^2\sqrt{\ln2})$.  The SPDC
process considered occurs in a $L=1$ mm lithium iodate ($LiIO_3$)
type I crystal, illuminated by a pump beam with $\lambda_p^0=405$
nm and a beam waist $\mathrm{w}_p=400$ $\mu$m. The generated
signal and idler photons are emitted with a wavelength
$\lambda_{s}^0=\lambda_{i}^0=810$ nm at an angle
$\varphi=\varphi_{s}=\varphi_{i}=10^{\circ}$. We consider
negligible Poynting vector walk-off ($\rho_0=0$) and in this case
all the values of $\alpha$ are equivalent.
\par

In figure ~\ref{wsdeltas} we see that the purity of the spatial
two photon state can be different from $1$, which implies a
correlation
 between space and frequency.  The strength of the correlation
 increases as the purity decreases.  From this figure we can see that as
 $\mathrm{w}_{s}$ and  $\mathrm{w}_{i}$
 increase the spatial purity of the two-photon state gets closer
 to $1$.  This can be understood intuitively since an infinitely big
 $\mathrm{w}_{s}$ ($\mathrm{w}_{i}$) implies the case of collecting only one
 $\mathbf{q}_{s}$ ( $\mathbf{q}_{i}$) and therefore the two-photon state is
separable in frequency and space.

The separability and the lack of correlation between frequency and
space, can also be seen in the ``narrow band'' limit where
$\Delta\lambda_{s}=\Delta\lambda_{i}\rightarrow0$
 nm: for any value of  $\mathrm{w}_{s}=\mathrm{w}_{i}$ the purity is always
 $1$.  For a typical commercial interference filter, $\Delta\lambda_{s}=\Delta\lambda_{i}\approx1$ nm, the correlation
between space and frequency is  significant and consequently the
purity decreases as can be observed in the figure. As the width of
the frequency filters increases,  the purity converge quickly
enough to make the case of
$\Delta\lambda_{s}=\Delta\lambda_{i}=10$ nm almost equivalent to
the case without frequency filters
$\Delta\lambda_{s},\Delta\lambda_{i}\rightarrow\infty$.

\begin{figure}[thb]
\centering
\includegraphics[bb= 4   326   593   700, scale=0.70]{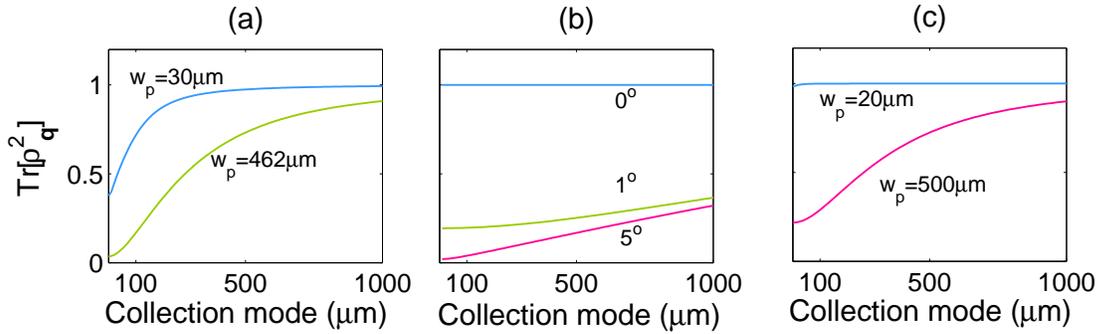}
\caption{Spatial purity as function of the collection mode for
three experimental cases.  In (a) is showed the case of reference
\cite{valencia}, in (b) the case described in reference
\cite{teich1} and in (c) the case reported in reference
\cite{altman}.} \label{comparation1}
\end{figure}

In order to compare these theoretical results with previous works,
in figure 3 we show  the purity of the spatial state as a function
of the collection mode, for three reported experiments. In figure
\ref{comparation1} (a) we used the data of reference
\cite{valencia}. In their case, a type I SPDC process in a $1$mm
long $LiIO_3$ crystal is analyzed. As a pump beam their considered
a diode laser with wavelength $\lambda_{p}=405$nm and bandwidth
$\Delta\lambda_p=0.4$nm. The pairs of photons generated propagate
at $\varphi=17^o$ and are detected by using monochromators with
$\Delta\lambda_{s,i}=0.2$nm. Two cases are depicted, the first one
when the pump beam is focalize with a waist of $w_p=30\mu$m over
the crystal and the second one where the pump beam waist is
$w_p=462\mu$m.  Their main result is the control of the frequency
correlations by changing the spatial properties of the pump beam.
As the pump beam influence the spatial shape of the generated
photons, they report implicitly a correlation between the space
and the frequency. For their spatial collection mode
$w_s=133,48\mu$m it can be seen in the figure that according with
our theoretical model the correlation appears.

Figure 3(b) shows $Tr[\rho_{{\bf q}}^2]$ for  the case described
in reference \cite{teich1}. A continues wave pump beam with
$\lambda_{p}=405$nm is used to generate photons  with
$\lambda_{s,i}=810$nm in a $1.5$mm length BBO crystal.
Interference filters of $\Delta\lambda_{s,i}=10$nm are used before
the detection. The pump beam waist satisfies the condition
$w_p>>L$ and the generated photons propagates at $\varphi=0^o$. As
it can be seen from the plot, the lack of correlation in this case
is due to the collinear configuration. Under the conditions
reported, but in a non collinear configuration will not be
possible to neglect the space and frequency correlations.

Finally in figure 3(c), $Tr[\rho_{{\bf q}}^2]$ is plotted in the
case described in reference \cite{altman}. In their case a pump
beam with $\lambda_p=351.1$nm is focused with a waist of
$w_p\approx20\mu m$ on a $2$mm long BBO crystal. Generated photons
at $\lambda_{s,i}=702$nm, propagating at $4^o$, are collected
after passing  $10$nm interference filters. In this noncollinear
case, the lack of correlations are due the small pump beam waist.
As can be seen in the figure, for less focused pump beams the
correlations become appreciable.

\section{Spatiotemporal entanglement between signal and idler}\label{svsi}
The correlations between frequency and space in the two photon
states showed in the previous section, suggest that for a complete
description of the entanglement between the photons,  both degrees
of freedom need to be taken into account.

In contrast with the previous section, here the physical system
considered is composed of two photons, each one described with
space and time variables, like is shown in figure
~\ref{comparation} (b). In what follows, we will calculate the
purity of the signal photon. Since the global state given by
equation~(\ref{densitymatrix}) is pure, the result of this section
allows us to calculate the degree of spatiotemporal entanglement
between signal and idler.

The reduced density matrix in space and frequency for the signal,
calculated by making a partial trace over the idler photon, writes

\begin{eqnarray}
    \rho_{signal} &=& Tr_{idler}(\rho)\\\nonumber
    &=& \int d \mathbf{q}_i''d\Omega_i'' \langle \mathbf{q}_i'',\Omega_i''|\rho|\mathbf{q}_i'',\Omega_i''\rangle\\\nonumber
    &=&\int d\mathbf{q}_sd\Omega_sd\mathbf{q}_id\Omega_i
    d\mathbf{q}_s'd\Omega_s'\\\nonumber
    &&\Phi(\mathbf{q}_s,\Omega_s,\mathbf{q}_i,\Omega_i)
    \Phi^{*}(\mathbf{q}'_s,\Omega_s',\mathbf{q}_i,\Omega_i)
    |\mathbf{q}_s,\Omega_s\rangle\langle
    \mathbf{q}'_s,\Omega_s'| \\\nonumber
\label{signalrho}
\end{eqnarray}

\noindent and its purity is given by

\begin{eqnarray}     \label{hola}
    Tr(\rho_{signal}^2)&&=
    \int d\mathbf{q}_sd\Omega_sd\mathbf{q}_id\Omega_i
    d\mathbf{q}_s'd\Omega_s'd\mathbf{q}_i'd\Omega_i'\\\nonumber
    &&\times\Phi(\mathbf{q}_s,\Omega_s,\mathbf{q}_i,\Omega_i)
    \Phi^{*}(\mathbf{q}'_s,\Omega_s',\mathbf{q}_i,\Omega_i)\\\nonumber
    &&\times\Phi(\mathbf{q}_s',\Omega_s',\mathbf{q}_i',\Omega_i')
    \Phi^{*}(\mathbf{q}_s,\Omega_s,\mathbf{q}_i',\Omega_i').
     \\\nonumber
\end{eqnarray}

\noindent Recalling the exponential character of the mode function
described by equation ~(\ref{exponentialMF}), we can write
equation ~(\ref{hola}) as
\begin{equation}\label{puritysignal}
    Tr(\rho_{signal}^2)=\frac{\det(2A)}{\sqrt{\det(C)}},
\end{equation}

\noindent where $C$ is a positive-definite real $12\times12$
matrix given by

\begin{eqnarray}\label{exponentialPS}
&&N^4\exp
    \left[-\frac{1}{2}X^tC X\right]=
\Phi(\mathbf{q}_s,\Omega_s,\mathbf{q}_i,\Omega_i)\\\nonumber
    &&
   \times \Phi^{*}(\mathbf{q}'_s,\Omega_s',\mathbf{q}_i,\Omega_i)
   \Phi(\mathbf{q}_s',\Omega_s',\mathbf{q}_i',\Omega_i')
    \Phi^{*}(\mathbf{q}_s,\Omega_s,\mathbf{q}_i',\Omega_i').
\end{eqnarray}

\noindent Notice that the difference between the matrix $C$ and
matrix $B$, in the last section, is the order of the primed and
unprimed variables in the arguments of the mode function.
\begin{figure}[thb]
\centering
\includegraphics[bb= 62   224   530   670,scale=0.50]{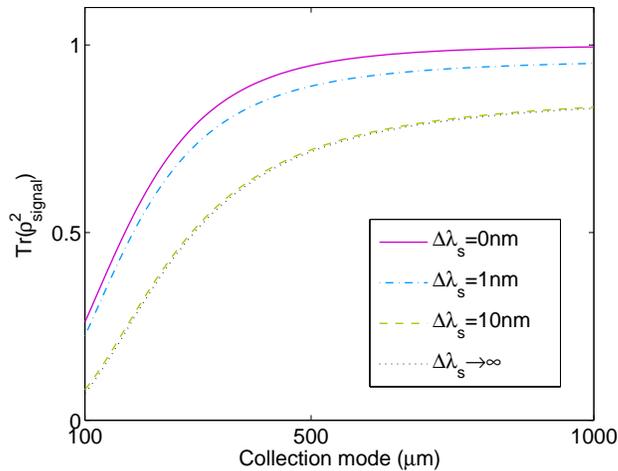}
\caption{Purity of the spatial frequency  state of the signal
photon for different values of the frequency filter bandwidth as a
function of the spatial collection mode width
($\mathrm{w}_{s}=\mathrm{w}_{i}$). The pump beam waist is
$\mathrm{w}_p=400\mu m$ and the emission angle is
$\varphi=10^{\circ}$.} \label{signalfilters}
\end{figure}

Figure  \ref{signalfilters} shows the space-frequency purity of
the signal photon as a function of the spatial filter width for
different values of the frequency filter width.  Maximal
separability for the photons is achieved when infinitely narrow
filters in space and frequency are used. In the region of small
values of $\mathrm{w}_{s}$ and $\mathrm{w}_{i}$,  the correlations
between frequency and space are considerable  even in the case of
infinitely narrow frequency filters. Different values for  the
signal photon purity can be achieved by changing the filters
width, as it can be seen from figure~\ref{signalfilters}. It is
important to take into account that $Tr(\rho_{signal}^2)$ is
confined between the values obtained for
$\Delta\lambda_{s}=\Delta\lambda_{i}\rightarrow0$ nm and
$\Delta\lambda_{s},\Delta\lambda_{i}\rightarrow\infty$. This
limits can be tailored by modifying the  SPDC configuration.
Furthermore, in specific configurations where non infinite filter
are used in space and/or frequency, the purity can be tailored by
using other parameters.

\begin{figure}[thb]
\centering
\includegraphics[bb=1   335   584   538,scale=0.70]{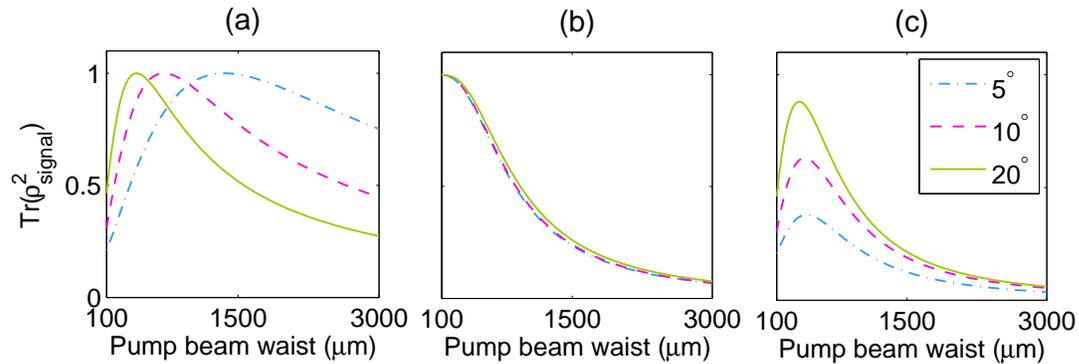}
\caption{Spatiotemporal purity of the signal photon for different
values of the emission angle, as a function of pump beam width
($\mathrm{w}_p$). In (a) $\Delta\lambda_{s}=\Delta\lambda_{i}=10$
nm and
 $\mathrm{w}_{s},\mathrm{w}_{i}\rightarrow\infty$.  In (b)
$\Delta\lambda_s,\Delta\lambda_i\rightarrow 0$ nm and
$\mathrm{w}_{s}=\mathrm{w}_{i}=400$ $\mu$m. In (c)
$\Delta\lambda_s=\Delta\lambda_i=10$ nm and
$\mathrm{w}_{s}=\mathrm{w}_{i}=400$ $\mu$m.}
 \label{angles}
\end{figure}

Figure  \ref{angles} shows $Tr(\rho_{signal}^2)$ as a function of
the pump beam waist $\mathrm{w}_p$ for various values of the
emission angle $\varphi=\varphi_{s}=\varphi_{i}$. In figure
~\ref{angles} (a) we considered a frequency filter
$\Delta\lambda_{s}=\Delta\lambda_{i}=10$ nm with infinitely narrow
spatial filters. In this case,  maximal purity can be found for
each emission angle at a particular value of the pump beam waist.
For a highly narrow band frequency filter  with a
$\mathrm{w}_{s}=\mathrm{w}_{i}=400$ $\mu$m spatial filter, figure
~\ref{angles} (b) shows how the correlation between the photons is
minimal for small pump beams being irrelevant the emission angle,
for this particular crystal length. Finally, in
figure~\ref{angles} (c) the case of finite spatial and frequency
filters is despicted. As can be seen, the purity of the signal
photon is always smaller than one, it increases for noncollinear
angles and it has a maximum for a given value of the pump beam
waist.

From the purity, equation~(\ref{puritysignal}), the amount of
entanglement between the signal and idler photons can be
calculated by using,  for example, the I-concurrence
$C=\sqrt{2(1-Tr(\rho_{signal}^2))}$ \cite{walborn,mintert} or the
Schmidt number $K=1/Tr(\rho_{signal}^2) $ \cite{vanexter2}.\\\par

\section{Conclusion}

An analytical expression that fully characterizes the purity of
the spatial two-photon state was obtained as a function of the
parameters involved in the SPDC process, equation
 (\ref{purityfreq}). The fact that the spatial two-photon state may
be mixed reveals a correlation between space and frequency that
cannot be neglected even in the cases of very narrow filtering.
This results were compared to previous analysis where the
correlations between space and frequency are neglected
\cite{teich1, altman} and with other one in witch the correlations
appears \cite{valencia}. This correlation is mathematically
analogous to the entanglement in the sense that the mode function
$\Phi(\mathbf{q}_s,\Omega_s,\mathbf{q}_i,\Omega_i) \neq
\Phi_{\mathbf{q}}(\mathbf{q}_s,\mathbf{q}_i)
\Phi_{\Omega}(\Omega_s,\Omega_i)$.  However, the concept of
entanglement is strongly associated to the possibility of a
spatial separation of the correlated subsystems.

The purity of the signal photon in space and frequency has been
also calculated, equation~(\ref{puritysignal}). This expression
represents a tool to tailor the space and frequency purity of the
signal photon by means of the geometrical SPDC configuration and
the filtering process.

\ack We thank M. Navascues for helpful discussions. This work was
supported by projects FIS2007-60179 and Consolider-Ingenio 2010
QOIT from Spain, by the European Commission (Qubit Applications
(QAP), Contract No. 015848) and by the Generalitat de
Catalunya.\par

\end{document}